\documentclass[runningheads]{llncs}
\usepackage{graphicx}
\usepackage{xcolor}
\usepackage{enumitem}
\setlist{nolistsep}
\usepackage{xspace}
\newcommand{\etal}{\emph{et al. }}
\newcommand{\ie}{\emph{i.e.}, }

\newcommand{\ECAS}{\texttt{ECAS-ML}\xspace}

\usepackage{algorithmic}
\usepackage[linesnumbered, ruled, vlined]{algorithm2e}

\begin{document}
\title{ECAS-ML: Edge Computing Assisted Adaptation Scheme with Machine Learning for HTTP Adaptive Streaming}

\titlerunning{ECAS-ML: Edge Computing Assisted Adaptation Scheme with ML for HAS}
%
\author{Jesús Aguilar-Armijo\orcidID{0000-0002-9551-4842} \and
Ekrem \c{C}etinkaya\orcidID{0000-0002-6084-6249} \and
Christian Timmerer\orcidID{0000-0002-0031-5243}
\and
Hermann Hellwagner\orcidID{0000-0003-1114-2584}}
\authorrunning{J. Aguilar-Armijo, E. \c{C}etinkaya, et al.}
%
\institute{Christian Doppler Laboratory ATHENA, Institute of Information Technology, Alpen-Adria-Universität Klagenfurt, Austria\\
\email{\{firstname.lastname\}@aau.at}\\}

\maketitle

\begin{abstract}
As the video streaming traffic in mobile networks is increasing, improving the content delivery process becomes crucial, e.g., by utilizing edge computing support. At an edge node, we can deploy adaptive bitrate (ABR) algorithms with a better understanding of network behavior and access to radio and player metrics. In this work, we present \ECAS, Edge Assisted Adaptation Scheme for HTTP Adaptive Streaming with Machine Learning. \ECAS focuses on managing the tradeoff among bitrate, segment switches and stalls to achieve a higher quality of experience (QoE). For that purpose, we use machine learning techniques to analyze radio throughput traces and predict the best parameters of our algorithm to achieve better performance. The results show that \ECAS outperforms other client-based and edge-based ABR algorithms.

\end{abstract}

\section{Introduction}
\label{sec:Introduction}

Video streaming traffic today represents a significant fraction of mobile network traffic. Therefore, it became very important to assure a good QoE to the video clients.
HTTP Adaptive Streaming (HAS) became the \textit{de facto} standard for video streaming. HAS divides the content into chunks or segments, each one encoded in different qualities, which allows adapting to changing network conditions. The ABR algorithm decides which segment on which quality level to request. We can classify ABR algorithms into four main categories: \textit{(1)} client-based adaptation, \textit{(2)} server-based adaptation, \textit{(3)} network-assisted adaptation, and \textit{(4)} hybrid adaptation~\cite{ABRsurvey}.
The most popular category is client-based adaptation. It provides scalability because the ABR algorithm runs in each client device. However, client-based adaptation has the problem that a client is not aware of what other clients are requesting.

Edge computing brings storage and computing power closer to the clients~\cite{edgeakey}. At the edge computing node, we also have access to radio metrics using the Radio Network Information Service (RNIS)~\cite{RNIS} as well as to clients' player metrics~\cite{3GPP}. Hence, an edge-based adaptation scheme has more computing power, storage and valuable information to run the ABR algorithm and make better decisions than clients independently.

Recurrent neural networks (RNNs)~\cite{RNN} are the common neural network structures when it comes to working with sequential data. Their structure allows them to capture the temporal dependencies thanks to their internal memory system. Long short-term memory networks (LSTMs)~\cite{LSTM} are a special type of RNNs in which the memory is extended.



In this work, we use machine learning techniques in an edge-based ABR mechanism to improve the QoE by managing the tradeoff among bitrate, segment switches, and stalls according to the current radio network conditions. LSTM is deployed to predict parameters for this tradeoff based on radio conditions. Furthermore, we compare the results against different client-based and edge-based ABR algorithms in diverse radio scenarios.

The main contributions of this paper are as follows:
\begin{itemize}
	\item the \ECAS system, an on-the-fly edge-based adaptation scheme with machine learning;
	\item the \ECAS ABR adaptation algorithm that manages the tradeoff among bitrate, segment switches and stalls to improve the QoE;
	\item the consideration of the device's screen resolution in the ABR algorithm, as higher screen resolutions demand higher bitrate to maintain a good QoE;
	\item machine learning techniques (\textit{i.e.,} parameter prediction using LSTM) to improve managing the tradeoff mentioned above according to the current radio network conditions; 
	\item a comprehensive evaluation of the \ECAS performance, including a comparison with other state-of-the-art client-based and edge-based ABR algorithms.
\end{itemize}

The remainder of this paper is structured as follows. Section~\ref{sec:RelatedWork} discusses related work. In Section~\ref{sec:ECAS} we present the proposed \ECAS approach. Section~\ref{sec:ExperimentalSetup} introduces the experimental setup we created to evaluate \ECAS. The results are described in Section~\ref{sec:Results}. Finally, Section~\ref{sec:ConclusionsAndFutureWork} concludes the paper and outlines future work.

\section{Related Work}
\label{sec:RelatedWork}
Although most of the ABR algorithm proposals are client-based, we can also find edge-based solutions.

In~\cite{SABR}, Bhat~\etal leverage the information of network conditions available in a software-defined network (SDN) to provide assistance to the video streaming delivery process, improving the final QoE of the clients. The adaptation algorithms of the video streaming clients remain unmodified. Furthermore, a better selection of caching strategies can lead to higher cache hit rates and, in consequence, an improvement of the content delivery process and QoE. 

In~\cite{fajardo2015improving}, Fajardo~\etal  introduce a new element in the mobile network architecture called ME-DAF to support multimedia delivery. ME-DAF implements content awareness, client awareness, and network awareness using the capabilities of edge computing. However, it is unclear if this scheme can outperform other edge-based ABR algorithms.


Kim~\etal~\cite{kim2020edge} propose an Edge Computing Assisted Adaptive Streaming Scheme for Mobile Networks that focuses on optimizing QoE, fairness and resource utilization. Moreover, they design an optimization model and a greedy-based ABR algorithm. Their results outperform other existing edge-based solutions such as Prius~\cite{Prius}.

Aguilar-Armijo~\etal~\cite{EADAS} propose EADAS, an edge-based mechanism consisting of \textit{(i)} an adaptation algorithm and \textit{(ii)} a segment prefetching scheme that supports the client-based ABR algorithm by improving its decisions on-the-fly. EADAS leverages edge capabilities such as the availability of player metrics, radio metrics and all clients' requests, as well as storage and computing power to improve the final QoE and fairness of the video streaming clients, outperforming other ABR solutions. 

We can also find other work that combines edge-based approaches with machine learning techniques.

In~\cite{chang2019edge}, Chang~\etal introduce an edge-based adaptive scheme that uses Q-learning techniques to select the adequate bitrate during the video streaming session in shared networks. Moreover, they consider radio metrics provided by the RNIS. The proposed scheme is evaluated against client-based algorithms such as buffer-based, rate-based and dashJS, but not against edge-based ABR algorithms.

Ma~\etal~\cite{steward} present Steward: Smart Edge based Joint QoE Optimization for Adaptive Video Streaming. Their mechanism optimizes the QoE and fairness under bandwidth bottlenecks using an edge-based ABR algorithm based on neural networks and reinforcement learning. In our understanding, the comparison with other ABR algorithms should also include edge-based algorithms and should be made with different radio network conditions.

\section{The ECAS-ML System}
\label{sec:ECAS}

\subsection{System Architecture}
\label{sec:architecture}

\begin{figure}[h]
    \centerline{\includegraphics[width=0.9\linewidth]{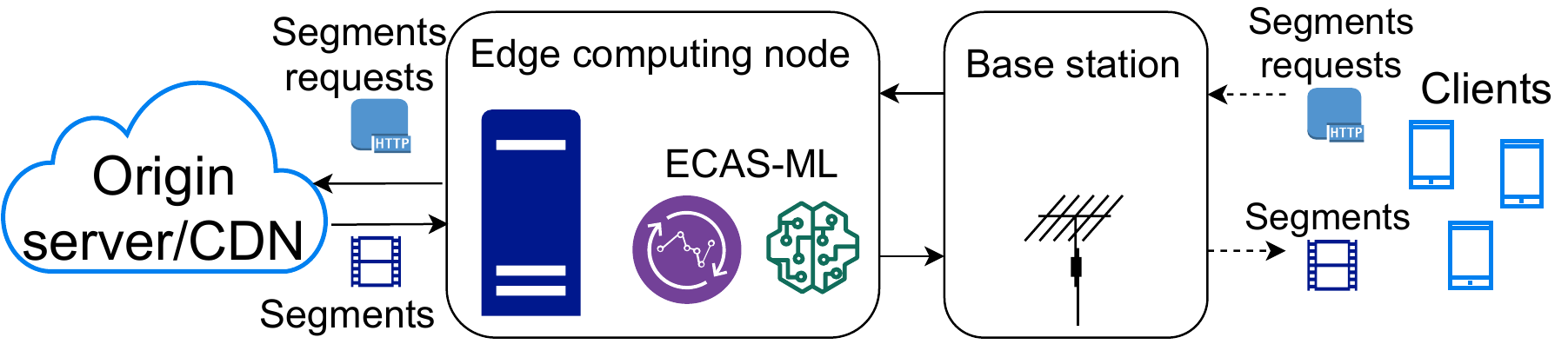}}
    \caption{\ECAS system architecture.}
    \label{fig:ecasarchtitecture}
\end{figure}

We propose an edge-based adaptation mechanism for HAS named \ECAS. The system architecture of \ECAS is shown in Figure~\ref{fig:ecasarchtitecture}. When a client sends a segment request to the video server, it is intercepted by the \ECAS mechanism located at the edge computing node. Next, the adaptation algorithm is executed on-the-fly and sends the modified segment request to the server. During the whole video streaming session, \ECAS requests radio information from the RNIS and machine learning techniques are used to  predict the best set of parameters by analyzing the radio traces periodically to provide a better adaptation. When the segment is served, the edge computing node forwards it to the client.

\ECAS is designed to be located on an edge node close to a base station in a cellular network. At this location, we have access to the necessary storage and computing power for running machine learning techniques.
Moreover, we have access to radio metrics provided by the RNIS and to player metrics such as buffer size that are reported periodically by the clients using the HTTP POST protocol. This process was standardized by 3GPP~\cite{3GPP}.


\subsection{ECAS Algorithm}
\label{sec:algorithm}
There are four main factors that affect the QoE: bitrate, segment switches, stalls and screen resolution. There is a tradeoff among bitrate, segment switches and stalls where the improvement of one metric may degrade the others. For example, if we want to achieve a higher bitrate, it might be at the risk of possible stall events in case the radio throughput experiences fading.

Another factor that affects the QoE is the screen resolution of the device that is playing the video. The same bitrate leads to different QoE perceived by the user, \ie higher resolutions require higher bitrates to maintain a good QoE. This fourth factor is often ignored in state-of-the-art ABR algorithms. \ECAS considers the screen resolution in the QoE measurement as nowadays different devices with different resolutions are connected to the mobile networks consuming video streaming content.

\begin{table}[ht]
\vspace{-5mm}
\centering
\caption{Notation used in this paper.}
\label{tab:notation}
\resizebox{0.95\linewidth}{!}{
\begin{tabular}{c|l}
\hline
\textbf{Symbol}  & \textbf{Definition} \\ 
\hline
{$Q$} & Number of quality representations \\
{$N$} & Window size (number of segments) \\
{$L$} & Segment length (sec.) \\
{$r_{s,t}$} & Bitrate of the quality s of the segment number t (kbps)\\ 
{$r'_{s,t}$} & Bitrate score of the quality s of the segment number t (kbps)\\
{$\overline{r_{t..t-N}}$} & Mean bitrate of the last N+1 segments (kbps)\\
{$B$} & Current buffer level of the current client (sec.)\\
{$B_{s,t+1}$} & Predicted buffer level after requesting quality s of the segment number t+1 (sec.)\\
{$dt_{s,t+1}$} & Download time of quality s of the segment number t+1 (sec.)\\
\hline
\end{tabular}
}
\end{table}

\resizebox{\columnwidth}{!}{
\begin{algorithm}[H]
 \vspace{5pt}
 // This algorithm is executed for each segment request for each client
 
    \KwData{$switches\_penalty\_factor$, $stalls\_penalty\_factor$, $buffer\_threshold\_1$, $buffer\_threshold\_2$, $screen\_resolution$, $est\_throughput$, $Q$, $N$, $L$, $B$, $\overline{r_{t..t-N}}$}
    \KwResult{Quality index to request ($quality\_to\_request$)}
    $QoE\_score = 0$\;
    $quality\_to\_request = 0$\;
    $best\_score = 0$\;
    
    // For each quality index \\
    \For{s = 0, 1, 2, ... Q-1}{
        \If{$screen\_resolution == "240p"$}{
            $\beta = 8.17$\;
        }
        \If{$screen\_resolution == "360p"$}{
            $\beta = 3.73$\;
        }
        \If{$screen\_resolution == "480p"$}{
            $\beta = 2.75$\;
        }
        \If{$screen\_resolution == "720p"$}{
            $\beta = 1.89$\;
        }
        \If{$screen\_resolution == "1080p"$}{
            $\beta = 0.78$\;
        }
        \If{$screen\_resolution == "2160p"$}{
            $\beta = 0.5$\;
        }
        $r'_{s,t+1} = r_{s,t+1} \times (1 - e^{-\beta \times r_{s,t+1} \times 0.001}) $
        
        $ \overline{r_{t+1..t-N}} = \frac{{(\overline{r_{t..t-N}} \times (N+1)) + r_{s,t+1}}}{N+2} $
        
        $switches\_penalty = \vert \overline{r_{t+1..t-N}} - r_{s,t+1} \vert \times switches\_penalty\_factor $ 
        
        $dt_{s,t+1} = \frac{r_{s,t+1} \times L}{est\_throughput}  $
        
        $B_{s,t+1} = B + L - dt_{s,t+1}$
        
        \eIf{$B_{s,t+1} < L \times buffer\_threshold\_1$}{
            // Buffer in high risk area, we do not consider that quality \\
             $QoE\_score = -inf$\;
        }{
            \eIf{$B_{s,t+1} < L \times buffer\_threshold\_2$}{
                // Buffer in medium risk area \\
                $b\_dif = L \times buffer\_threshold\_2 - B_{s,t+1}$
                $stalls\_penalty = b\_dif \times \overline{r_{t+1..t-N}} \times stalls\_penalty\_factor$
                $QoE\_score =  r'_{s,t+1} - switches\_penalty - stalls\_penalty$\; 
            }{ 
                // Buffer in low risk area \\
                 $QoE\_score =  r'_{s,t+1} - switches\_penalty $\; 
            }
        }
        \If{$s == 0$}{
             $best\_score = QoE\_score$\;
        }
        \If{$QoE\_score > best\_score$}{
             $best\_score = QoE\_score$\;
             $quality\_to\_request = s$\;
        }
 } 
\textbf{return} $quality\_to\_request$;\\

 \caption{\ECAS algorithm.}
 \label{alg}
\end{algorithm}
}

\begin{figure}[h]
    \centerline{\includegraphics[width=0.5\linewidth]{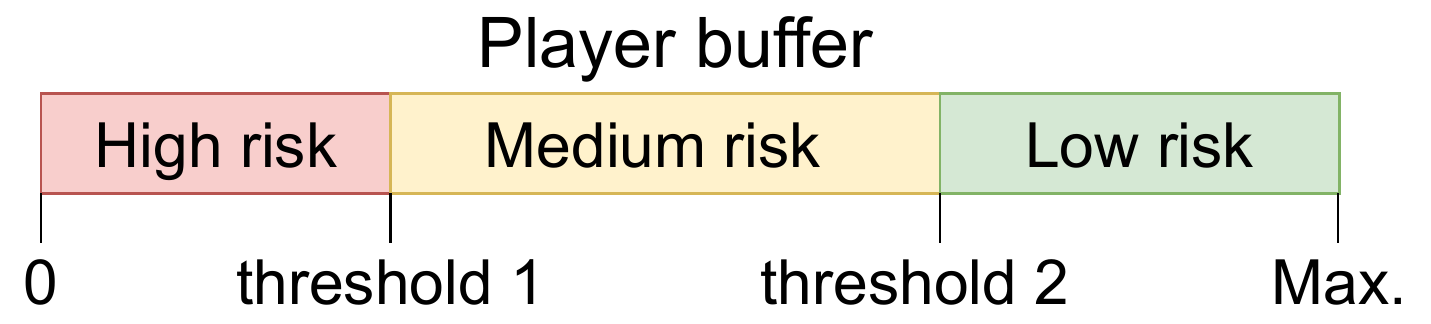}}
    \caption{\ECAS buffer areas.}
    \label{fig:buffer}
\end{figure}

We designed \ECAS to manage the tradeoff among bitrate, segment switches and stalls and to consider the screen resolution in order to maximize the final QoE perceived by the user. For that purpose, we introduce four variables in our algorithm:
\begin{itemize}
    \item \textit{Switches penalty factor}, an integer number that controls the segment switches. A high switches penalty factor indicates we prioritize reducing segment switches during the video streaming session.
    \item \textit{Stalls penalty factor}, an integer number that controls the stalls. The higher the stalls penalty factor, the fewer stalls will occur during the video streaming session.
    \item \textit{Buffer threshold 1}, an integer number that delimits the high-risk area of the buffer. A long high-risk area means more conservative segment requests, therefore, fewer stalls and lower mean bitrate. 
    \item \textit{Buffer threshold 2}, an integer number that delimits the low-risk area of the buffer. \ECAS focuses on maintaining the buffer in the medium-risk and low-risk areas, in order to keep the video streaming session stable. The different buffer areas are shown in Figure~\ref{fig:buffer}.
\end{itemize}

For each segment request of each user, Algorithm~\ref{alg} is executed. The algorithm evaluates all the different qualities and assigns each one of them a QoE score. Finally, it would request the segment quality with the highest QoE score.

For each segment quality, the process to assign a QoE score is the following:
\begin{itemize}
    \item \textbf{Phase 1} \textit{(lines 7 to 19)}: First, it considers the device's screen resolution to calculate the bitrate score $r'_{s,t+1}$. We use the equations and $\beta$ values shown in~\cite{ScreenResolution} as a baseline for our algorithm. The beta value models the curve that relates bitrate and Mean Opinion Score (MOS). We obtain and include beta values for 1080p and 2160p resolutions according to our experiments.
    \item \textbf{Phase 2} \textit{(lines 20 to 23)}: This phase of the algorithm calculates: \textit{(1)} the mean bitrate of the last $N+1$ segments plus the segment quality we are evaluating ($\overline{r_{t+1..t-N}}$); \textit{(2)} the switches penalty; \textit{(3)} the download time for the segment quality we are evaluating using the estimated throughput based on the radio and backhaul throughput available at the RNIS ($dt_{s,t+1}$); and \textit{(4)} the estimated player buffer size after receiving the quality we are evaluating ($B_{s,t+1}$).
    \item \textbf{Phase 3} \textit{(lines 24 to 33)}: In this phase, different lines of code are executed depending on the estimated buffer size and the buffer thresholds. We differentiate three risk areas: \textit{High risk:} there is a high risk of a stall if we request this quality, so its QoE score is set to the minimum which means we do not consider that quality. \textit{Medium risk:} there is a medium risk of a stall in future requests, so we apply a switches penalty and a stall penalty proportionally to the difference between the estimated buffer size and threshold 2. \textit{Low risk:} we consider there is no risk of stalls in the following segment request and only the switches penalty is applied.
    \item \textbf{Phase 4} \textit{(lines 34 to 38)}: Finally, if the QoE score of the quality we are evaluating is better than the previous best QoE score, we update the quality to request and the best score to beat.
\end{itemize}

\subsection{Parameter Prediction with Machine Learning}
\label{sec:ml}

The machine learning part of \ECAS is designed to process sequential data since radio traces consist of the throughput over time. We model this problem as predicting the most suitable parameters for the given radio trace, thus a regression problem.

RNNs are known to work well with sequential data. One downside of RNNs is short-term memory. If the input sequence is long, RNNs usually fail to utilize the early stage information to later stages. To address this issue LSTM~\cite{LSTM} is proposed. 

The memory structure in LSTM consists of three gates (\textit{i.e.,} input, forget, and output). When new input arrives, these gates can be used to perform three different actions: \textit{(1)} use the incoming information (input gate); \textit{(2)} delete the information (forget gate); and \textit{(3)} use the incoming information to impact the output (output gate).  This structure allows LSTM to remember the input for a longer time; thus, enabling the LSTM to exploit the dependencies when the temporal delay is higher. 


RNNs have one common drawback, which is the exploding gradients problem, and we used gradient clipping~\cite{GradientClip} to address this issue. Moreover, Huber loss~\cite{HuberLoss} with $\Delta = 1$ is used as the loss function as it is known to work better against exploding gradients compared to mean absolute error (MAE). Also, it combines the advantages of both mean squared error (MSE) and MAE losses. 

\begin{figure}[t]
    \centering
    \includegraphics[width=\linewidth]{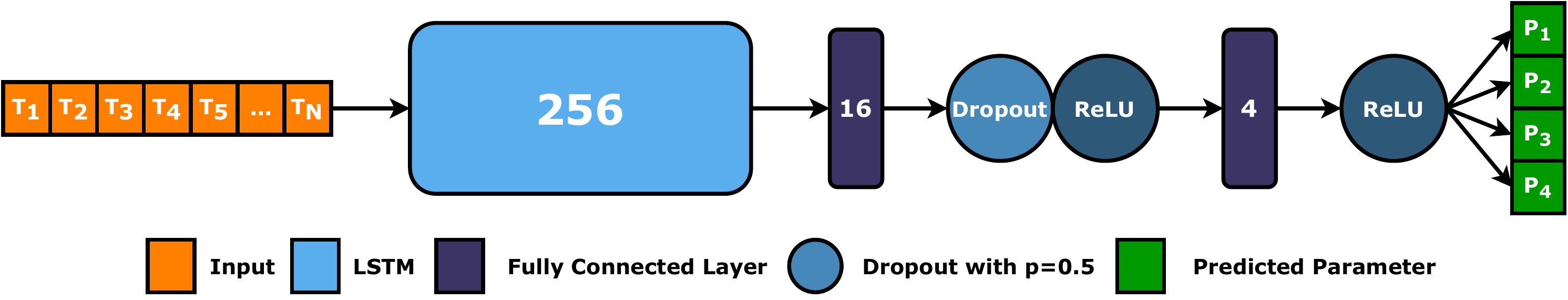}
    \caption{Structure of the proposed LSTM-based model. The numbers inside the boxes indicate the output feature size. $T_{N}$ is the throughput value at the $N$th second in the radio trace.}
    \label{fig:lstm}
\end{figure}

{We use an LSTM-based approach to predict parameters since LSTMs are known to exploit long-term dependencies effectively~\cite{GRUvsLSTM}. The proposed structure is illustrated in Fig.~\ref{fig:lstm}.
The proposed model takes the throughput over time as an input and predicts the optimal set of parameters for the given input vector. The input vector length increases as the streaming session continues and new parameters are predicted periodically.}

The ML model takes a radio trace as input which is the throughput per second, and predicts a vector with four values (\textit{i.e., switch penalty, stall penalty, threshold 1,} and \textit{threshold 2}). The input vector ($V_{t, s}$) definition for a given trace ($t$) and second ($s$) is given in Eq.~\ref{eq:input}.

\begin{equation}
\label{eq:input}
    V_{t, s} = [T_{t, 1}, T_{t, 2}, T_{t, 3}, ..., T_{t, s}]
\end{equation}

where ($T_{t, s}$) is the throughput in second $s$ for the radio trace $t$. Each value in the input vector is the throughput per second from the beginning of the streaming session until the current second. By following this approach, it is possible to utilize the ML model in the very early stages of the streaming session as the minimum length of the input vector is five (\textit{i.e.,} five seconds of the radio trace). Moreover, as the streaming continues, the ML model has more reliable data to predict, thus resulting in a better set of parameters.

Since the radio trace lengths can vary, the input size of the model should be adaptive. The ML model in \ECAS is designed to work with variable length traces. One common approach to follow in those situations is padding the input to match the maximum sequence length in the dataset. However, using zero padding is not suitable for our use case as the throughput can actually be zero from time to time. Thus, we did not apply any padding; instead, we trained the network with a single radio trace (\ie batch size 1) at a time.

\section{Experimental Setup}
\label{sec:ExperimentalSetup}

To test our proposed scheme \ECAS, we developed a Python-based edge computing and video streaming simulator that supports edge mechanisms, real radio traces and video streaming datasets. Its architecture consists of a video streaming server, an edge computing node, a base station and multiple clients. We consider the latency and the throughput between each path of the network to simulate the content delivery. The testbed follows the procedure explained in Section~\ref{sec:architecture}. Periodically, the edge computing node collects player and radio information to perform the adaptation decisions.

To compare performance, we implement three client-based ABR algorithms that follow three different approaches: throughput-based ABR (TBA~\cite{TBA}), buffer-based ABR (BBA~\cite{BBA}), and hybrid-based ABR (SARA~\cite{SARA}). Moreover, we implement three edge-based ABR algorithms: Greedy-Based Bitrate Allocation (GBBA)~\cite{kim2020edge}, EADAS~\cite{EADAS}, and our proposed scheme \ECAS.

We use the \textit{Big Buck Bunny}\footnote{\url{http://ftp.itec.aau.at/datasets/DASHDataset2014/BigBuckBunny/2sec/}} video from the streaming dataset~\cite{dataset}; it must be noted that similar results were obtained with other videos. We choose two second segments and bitrate levels of [50, 100, 150, 200, 250, 300, 400, 500, 600, 700, 900, 1200, 1500, 2000, 2500, 3000, 4000, 5000, 6000, 8000] kbps in resolutions ranging from ($320 \times 240$) to ($1920 \times 1080$). Half of the clients use a device with a 1080p screen resolution, and the other half with a 2160p screen resolution.

Our testbed provides different metrics such as mean bitrate, mean switching magnitude (measured in kbps and in quality indices), number of stalls, mean stall duration and the QoE according to the recommendation ITU-T P.1203~\cite{P1203_B}. We choose this QoE model as it is the first standardized audiovisual quality model for HAS and it has been widely trained and validated. 

For the radio traces, our simulator uses real radio traces from a 4G dataset~\cite{beyond4G} with different mobility patterns. This dataset consists of 5 trace categories, and each category contains a different number of radio traces (\textit{i.e., bus (16), car (53), pedestrian (31), static (15),} and \textit{train (20)}). There are 135 radio traces in total, and we removed 6 from the dataset since they were causing an imbalance in the dataset due to being too long. In the end, we used 121 traces for training and 8 for testing. Since each category has a different number of traces, we picked one trace from \textit{bus} and \textit{static} and two from \textit{car, pedestrian,} and \textit{train} for testing. 

For each radio trace in the training set, input vectors are extracted from the fifth second until the end of the trace. These vectors are then shuffled randomly and used for training the model. We used $136,466$ input vectors in total for training and $10,015$ for testing. We needed the optimal set of parameters as labels for input vectors in the training dataset. These parameters were found by applying a brute force approach in the simulation.

Pytorch~\cite{Pytorch} is used as the machine learning framework. The LSTM model is trained with Adam~\cite{Adam} as the optimizer with the learning rate of $5e-6$, and Huber loss~\cite{HuberLoss} with $\Delta = 1$ as the loss function. Dropout~\cite{Dropout} is applied after each layer to prevent overfitting. The \ECAS model is trained for 50 epochs with a gradient clipping applied to prevent the exploding gradients problem.

\section{Results}
\label{sec:Results}

\subsection{ECAS Performance Evaluation}
\label{sec:ECASperformance}

We compare the performance of \ECAS to other client-based and edge-based ABR algorithms as explained in Section~\ref{sec:ExperimentalSetup}.

\begin{table*}[h]
\centering
\caption{ECAS performance evaluation.}
\label{tab:ECAS}
\resizebox{1\textwidth}{!}{
\begin{tabular}{|l|c|c|c|c|c|c|c|}
\hline
\textbf{} & \textbf{BBA} & \textbf{TBA} & \textbf{SARA} & \textbf{GBBA} & \textbf{EADAS} & \textbf{ECAS-ML} \\ \hline
\textbf{Mean bitrate (kbps)} & 1314 & 2700 & 2700 & 2732 & 2668 & 2772 \\ \hline
\textbf{Mean switching magnitude (kbps)} & 325 & 921 & 633 & 991 & 958 & 1113 \\ \hline
\textbf{Mean switching magnitude (quality index)} & 1.02 & 4.00 & 3.14 & 3.32 & 4.14 & 4.21  \\ \hline
\textbf{Number of stalls} & 0 & 246 & 65 & 97 & 22 & 16 \\ \hline
\textbf{Mean stall duration (ms)} & 0 & 1107 & 1427 & 3153 & 2407 & 2523  \\ \hline
\textbf{QoE score (ITU-T P.1203 mode 0)} & 3.25 & 3.05 & 2.86 & 3.07 & 3.41 & 3.65 \\ \hline
\end{tabular}
}
\end{table*}

In Table~\ref{tab:ECAS} we show the mean metrics of the eight clients during the video streaming session. \ECAS has the highest mean switching magnitude, which decreases the QoE, but also the highest mean bitrate and a low number of stalls. The tradeoff among bitrate, segment switches and stalls is successfully managed as \ECAS achieves a high QoE, outperforming other client-based and edge-based algorithms. More concretely, \ECAS improves over the QoE of \textit{BBA} by 12.31\%, of \textit{TBA} by 19.67\%, of \textit{SARA} by 27.6,2\%, of \textit{GBBA} by 18.89\%, and of \textit{EADAS} by 7.04\%.

\textit{TBA}, \textit{SARA} and \textit{GBBA} have a high number of stalls since their algorithms are not conservative enough for sudden drops in the radio throughput as they occur in the real 4G traces we used to conduct these experiments. Even with a high mean bitrate, this high number of stalls decreases the final QoE.

\subsection{Buffer Behavior}
\label{sec:bufferBehavior}

We show the behavior of the buffer size over time in Figure~\ref{fig:buffer}. This simulation was made with one user with a car mobility pattern with the setup explained in Section~\ref{sec:ExperimentalSetup}.
The value of the first buffer threshold is 3 and the value of the second buffer threshold is 6, therefore, as the segment duration is 2 seconds, the buffer areas are as follows:
\begin{itemize}
    \item High-risk area: from 0 seconds to 6 seconds.
    \item Medium-risk area: from 6 seconds to 12 seconds.
    \item Low-risk area: from 12 seconds to 20 seconds (maximum buffer size).
\end{itemize}

\begin{figure}[ht]
    \centerline{\includegraphics[width=0.9\linewidth]{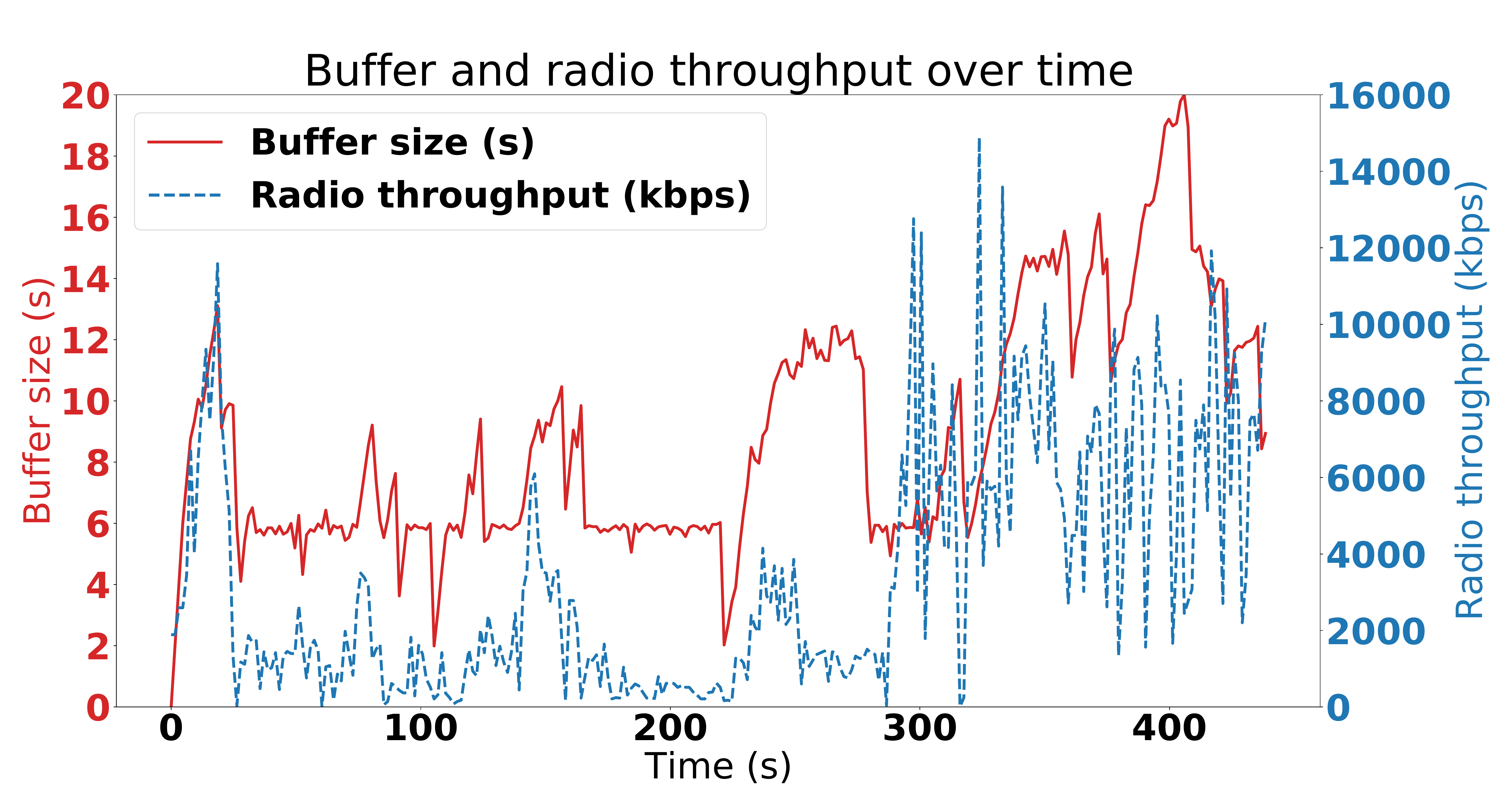}}
    \caption{Buffer size and radio throughput over time.}
    \label{fig:bufferover time}
\end{figure}

\ECAS provides the best quality possible while maintaining the buffer level above the first threshold, set at 6 seconds, and considering the segment switches penalty. Even in highly variable conditions such as the radio throughput traces used in this simulation, \ECAS prevents many stalls by avoiding the high-risk area where stalls occur more frequently. The second buffer threshold is not as critical as the first one, as it delimits the medium-risk and low-risk areas. If the buffer is in the low-risk area, it will not face any stalls penalty, just the switching penalty. Hence it would tend to request higher qualities as long as the radio throughput is high enough, and it will return to the medium-risk area if the radio throughput drops.

Determining proper values of both buffer thresholds is key for the adequate performance of \ECAS. If the first threshold is set too low, sudden radio throughput drops may lead to stall events that degrade the QoE. On the other hand, if the first threshold value is too high, the behavior would be too conservative, and we may not leverage all the network's resources.

\section{Conclusions and Future Work}
\label{sec:ConclusionsAndFutureWork}

In this work, we present \ECAS, an edge-based adaptation scheme for HAS. \ECAS focuses on achieving the best QoE managing the tradeoff among bitrate, segment switches and stalls. For this purpose, the \ECAS algorithm includes four variables: switches penalty factor, stalls penalty factor and two buffer thresholds. \ECAS also considers the device's resolution in its algorithm, as higher resolutions demand higher bitrates to achieve a good QoE. \ECAS utilizes an LSTM model to predict the optimal set of parameters for the given status of the radio trace.
Results show that \ECAS outperforms other ABR algorithms: client-based (BBA, TBA and SARA) and edge-based ones (GBBA and EADAS). In future work, the parameter prediction part of \ECAS can be improved using different ML techniques such as reinforcement learning.


\bibliographystyle{splncs04}
\bibliography{bibliography}

\end{document}